# Elastic Properties of Hot Isostatically Pressed Magnesium Diboride

V. F. Nesterenko[a] and Y. Gu

*Department of Mechanical and Aerospace Engineering*

*UCSD Materials Science and Engineering Program*

*University of California, San Diego*

*La Jolla, California 92093-0411*

**ABSTRACT**

Magnesium diboride was hot isostatically pressed using three qualitatively different cycles: cooling under pressure (DMCUP), "standard" cycle with pressure and temperature simultaneously reduced and isothermal pressure release. Elastic properties of dense $MgB_2$ were measured at normal conditions using resonant ultrasound spectroscopy (RUS) method. The highest values of elastic moduli (Young's modulus 273.5 GPa, bulk modulus 143.9 GPa, Poisson's ratio 0.183) correspond to the sample processed using DMCUP cycle. These data agree with theoretical predictions based on quantum mechanics calculations. Effect of lower density on elastic constants is consistent with theoretical approach based on elasticity theory taking into account effect of low porosity.

E-mail: vnestere@mae.ucsd.edu`

PACS indexes:  74.62Bf and 74.70.-b.

Discovery of superconductivity in magnesium diboride[1] at 39 K initiated intense research of properties of this new material. High density $MgB_2$ with sizes useful for some applications was fabricated at a pressure 0.2 GPa using DMCUP (dense material cooled under pressure) HIPing cycle[2,3] with excellent mechanical and superconducting properties.[2-8]

There is a lack of data on elastic properties of solid $MgB_2$.[9] This is mainly due to the fact that most methods (like powder-in-tube (PIT)[10]) result in a porous product. Recent quantum-mechanical calculations based on density functional theory[9] gave value for bulk modulus $K$ equal 139.3 GPa and 139 GPa[15] being consistent with other theoretical predictions[11] (140.1 GPa) and significantly different than theoretical results of[12,13] (163 GPa and 150 GPa). Data on X-ray powder diffraction in diamond anvil at $P \leq 8$ GPa gave value of $K = 120$ GPa[14] and 151 GPa.[15] Preliminary data for $K$ of HIPed samples[16] are consistent with theoretical predictions of.[9,11,15]

HIPing of encapsulated $MgB_2$ powder (–325 mesh size, 98% purity, Alfa Aesar, Inc.) was carried out using a three qualitatively different cycles (Figure 1). First one is a DMCUP cycle where the cooling of the densified material was performed under pressure (Figure 1(a)). This is helpful to reduce a local residual tensile stresses connected with temperature gradients and probable reaction between residual Mg and B on the stage of cooling, thermal mismatch between components and reduce microporosity due to the presence of Mg vapor at high temperatures. Similar cycle was successfully employed by Shields et al.[17] In a second cycle pressure and temperature are reduced simultaneously (Fig. 1(b)). The third cycle (Fig. 1(c)) utilizes isothermal pressure release which may be beneficial for some applications like HIPing of superconducting wires.[18] Description of the process can be found in.[16]

The quality of material after HIPing allowed preparation of high precision samples with accuracy of dimensions[16] suitable for resonant ultrasound spectroscopy (RUS).[19] Twelve samples of parallelepiped shape with dimensions from 1.6 to 13 mm were tested using DRS M$^3$odulus II system, supplied by DRS Inc. During machining one of IV-type sample was violently shattered into small pieces as a result of probable reaction of magnesium diboride with high density of dislocations[15] and water penetrating into pores. Samples of I-1 – I-3 with largest value of density and elastic constants were tested using two different sample holders (mm and cm type) with similar results.

Elastic properties in the RUS method are characterized based on the measurements of a large number of resonant frequencies with exceptionally high absolute accuracy.[19] Values of $c_{11}$ and $c_{44}$, estimates of error bars, number of resonant peaks and calculated elastic constants $E, K$ and $\nu$ are shown in Table 1. As a rule RUS method allows measurements of $c_{44}$ with higher accuracy than $c_{11}$.[19] A RMS deviations calculated by DRS RUS program were within a range of 0.029 – 0.050 percent for different samples and sample holders. Sensitivity of experimental data to the number of fitted frequencies can be illustrated by comparison of the reported data and the results for the same samples in.[16] Estimates of error bars were found using rpr.exe code developed by Migliori and Sarrao[19] and available in DRS M$^3$odulus II. Usually error bars are larger than RMS deviations.[19] RMS deviations calculated using rpr.exe code are few times larger than provided by DRS RUS program. Values of $c_{11}$ and $c_{44}$ calculated by DRS RUS program and by rpr.exe code (at fixed dimensions) were identical. Density was calculated using a mass and dimensions.

There is a correlation between the values of density, elastic moduli and $Q$. For example, $Q = 6.5 \bullet 10^3$ for the sample I-1 (Fig. 2(a)) with highest density and elastic moduli and $Q = 1.6 \bullet 10^3$ (I-4n, Fig.2(b)) and $Q = 1.9 \bullet 10^3$ (I-5n) for the samples with lowest density (in type-I samples) in the frequency range 1050 ÷ 1070 kHz. The lowest $Q = 4.2 \bullet 10^2$ in a similar frequency range was observed for samples IV (Figure 2(c)). As a rule, for a high density samples (I,II,III type) "Q-contrast" was typical − the first mode (n = 1), which depends only on $c_{44}$, has a lowest $Q$. For example in sample I-1 for mode n = 1. $Q = 3.3 \bullet 10^3$ and for the next 5 modes $Q$ is in the interval $(9 \div 6) \bullet 10^3$, less sharp "Q-contrast" was observed for samples II and III. Such behavior was not observed for samples IV with substantial porosity. At the same time $Q$ for the first mode (n = 1) for samples I-4n and IV-1 were practically the same (5.8 $\bullet 10^2$ and 5.5 $\bullet 10^2$ correspondingly) despite a few times larger $Q$ for higher modes in the sample I-4n. Values of $Q$ for the same resonance frequencies were dependent on sample holders (mm- or cm-stage) for samples I-1 – I-3 being of the same order of magnitude. Resonances depending mostly on $c_{44}$ (majority of them are of this type) have similar $Q$ as $c_{11}$ dominated modes.

There is a consistency between density and elastic constants suggesting that a porosity of material is a major reason for reduction of elastic properties. A variation of parameters is observed for a samples corresponding to the same HIPing run (I-1 ÷ I-3 and I-4*n* and I-5*n*). This may be caused by the variation of initial packing density or grain size affecting the consolidation process. It may indicate that plastic deformation of $MgB_2$ is not the main mechanism of consolidation and diffusion based mechanisms of densification are important.

Data for sample I-1 can be considered as representative of elastic constants of fully dense polycrystalline sample of $MgB_2$. The corresponding value of $K = 143.9$ GPa is between predicted values for bulk modulus (139, 139.3 and 140.1 GPa)[15,9,11], 150 GPa[13] and experimental value 151 GPa.[15] The highest values of elastic modules were measured in the samples processed by DMCUP cycle (Fig.1(a)).

There is a difference between properties of samples I and II taken from different HIPing runs at the same conditions for different size of HIPed cylinder. This raises a question of property dependence on the size of HIPed cylinder, which needs more detailed investigation.

The value of the Young's modulus at room temperature for $MgB_2$ prepared by reacting Mg and B in a sealed stainless tube (PIT method) was substantially lower (167 GPa)[21] than our data (Table 1) for HIPed material corresponding to lower density 2.27 g/cm$^3$ in PIT method.

We can evaluate expected difference in elastic modules due to porosity using theoretical results for medium with low density of spherical cavities with fraction $f$.[20]  Bulk modulus ($K_{av}$) for porous media ($K$ is a value for solid) is:[20]

$$(K_{av}/K)_{th} = 1 - 3f(1-v)/2(1-2v) + 0(f^2), \qquad f = (\rho_s - \rho_p)/\rho_s,$$

where $\rho_s$ and $\rho_p$ are densities of solid magnesium diboride and porous sample correspondingly.

Density of sample I-1 (2.66 g/cc), which is close to a reported value 2.625 g/cc based on X-ray measurements,[22] can be considered as a value for $\rho_s$ and $K$ = 143.9 GPa, $v$ = 0.183. Theoretical estimates for bulk modulus agree with experimental values within ten percent for samples with low porosity I-4n, I-5n, II and III.  It supports conclusion that observed difference in elastic moduli is due to the porosity of these samples.  At larger porosity (samples IV) theoretical value is larger than experimental value by about 25 percent.  The similar agreement between theoretical and experimental values at low porosity and overestimation of theoretical $K$ at larger porosity ($f$ = 0.26) can be also observed based on data for MgO.[23]

In summary, the highest bulk elastic modulus of HIPed $MgB_2$ is in a good agreement with theoretically predicted values based on quantum mechanics calculations.  Variations of elastic moduli with low porosity are in agreement with estimates based on continuum elasticity theory.

**Acknowledgements**

This work was supported by UCSD start-up fund of one of the authors. R. Muzyka (Flow Autoclave Systems, Inc) provided excellent support in extensive use of CIP and HIP machines at

## Figure captions

**Figure 1**. Three cycles employed to fabricate dense magnesium diboride: (a) DMCUP cycle (samples I, II), (b) "standard" cycle (samples III) and (c) cycle with isothermal pressure release (samples IV).

**Figure 2**. Resonant peaks for samples with different porosity I-1, fully dense sample (a), I-4n, $f = 0.068$ (b), and IV-2n, $f = 0.124$ (c). Data are obtained with mm-stage sample holder.

**Figure 1**

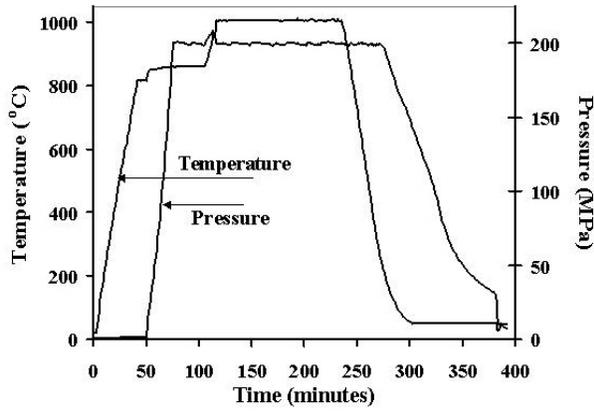

(a)

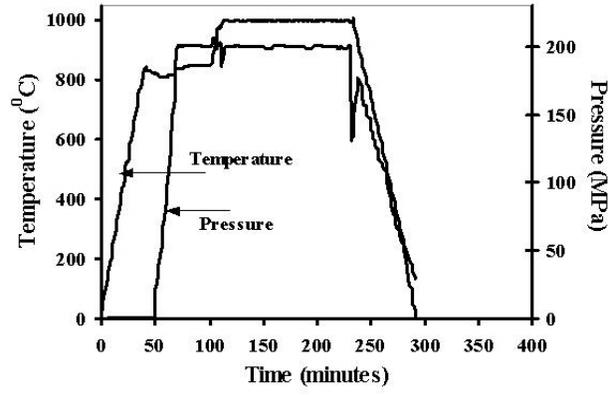

(b)

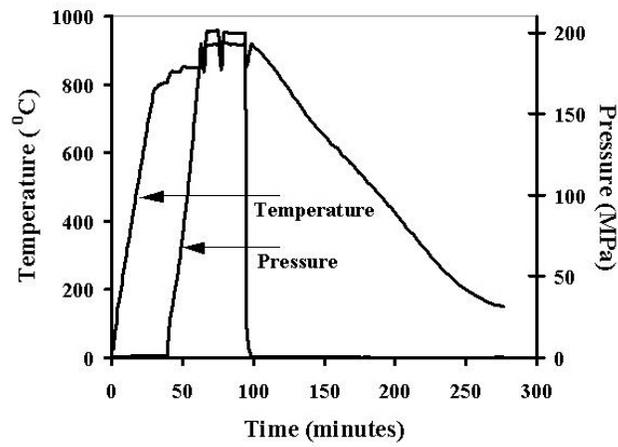

(c)

**Figure 2**

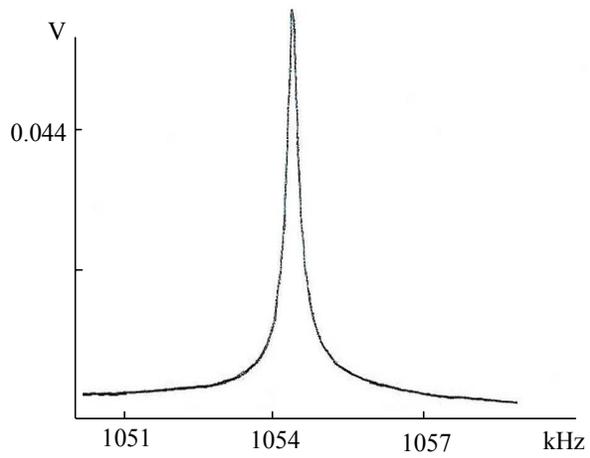

(a)

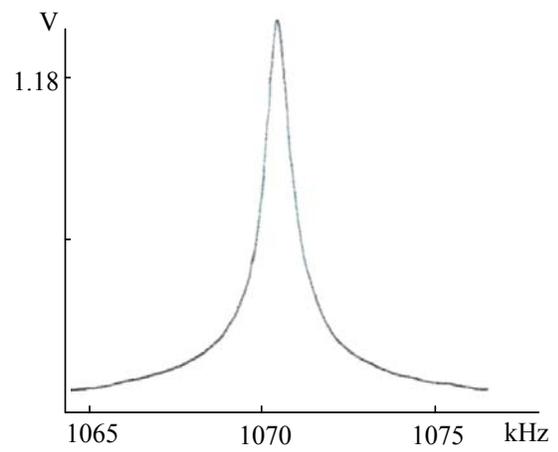

(b)

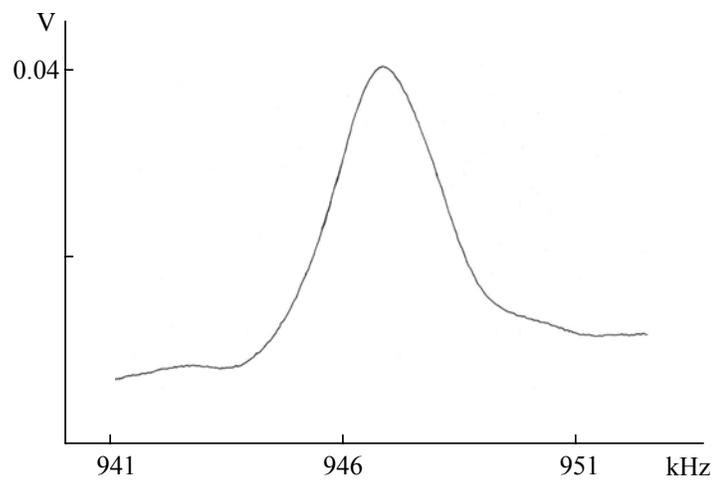

(c)

**Table 1**. Measured and calculated elastic properties of HIPed $MgB_2$.

| # Sample (type of stage) | $\rho$, g/cc | $C_{11}$, GPa | Error bar, % | $C_{44}$, GPa | Error bar, % | $E$, GPa | $K$, GPa | $v$ | Number of peaks |
|---|---|---|---|---|---|---|---|---|---|
| Sample I-1 (mm) | 2.66 | 298.0 | 0.17 | 115.57 | 0.05 | 273.5 | 143.9 | 0.183 | 42 |
| Sample I-2 (mm) | 2.63 | 297.7 | 0.30 | 114.78 | 0.07 | 272.3 | 144.7 | 0.186 | 47 |
| Sample I-3 (mm) | 2.63 | 298.3 | 0.39 | 114.58 | 0.10 | 272.3 | 145.5 | 0.188 | 46 |
| Sample I-4*n* (mm) | 2.48 | 246.3 | 0.2 | 96.47 | 0.05 | 227.3 | 117.7 | 0.178 | 47 |
| Sample I-5*n* (mm) | 2.48 | 246.4 | 0.12 | 96.21 | 0.03 | 227.0 | 118.1 | 0.180 | 47 |
| Sample II-1 (cm) | 2.56 | 266.7 | 0.16 | 103.79 | 0.04 | 245.2 | 128.3 | 0.181 | 43 |
| Sample II-2 (cm) | 2.56 | 266.3 | 0.22 | 104.10 | 0.06 | 245.5 | 127.5 | 0.179 | 46 |
| Sample II-3 (cm) | 2.56 | 265.5 | 0.10 | 103.75 | 0.03 | 244.7 | 127.2 | 0.179 | 41 |
| Sample III-1 (cm) | 2.51 | 249.6 | 0.09 | 97.47 | 0.03 | 230.0 | 119.6 | 0.180 | 50 |
| Sample III-2 (cm) | 2.53 | 251.3 | 0.13 | 98.38 | 0.04 | 231.8 | 120.1 | 0.178 | 39 |
| Sample IV-1 (mm) | 2.29 | 218.0 | 0.55 | 99.71 | 0.27 | 215.1 | 85.1 | 0.079 | 21 |
| Sample IV-2n (mm) | 2.33 | 212.4 | 1.33 | 94.48 | 0.48 | 207.7 | 86.4 | 0.099 | 14 |